\documentclass[prb,floats,twocolumn]{revtex4}

\usepackage{graphicx}
\usepackage{amsfonts}
\usepackage{amssymb}
\usepackage{amsbsy}
\usepackage{amsmath}

\begin{document}

\title{Magnetic phase diagram of the infinite-$U$ Hubbard model with nearest- and next nearest-neighbor hoppings}

\author{G. G. Blesio, M. G. Gonzalez, F. T. Lisandrini}

\affiliation {Instituto de F\'{\i}sica Rosario (CONICET),
Boulevard 27 de Febrero 210 bis, (2000) Rosario, Argentina} 

\date{\today}

\begin{abstract} 
We study the infinite-$U$ Hubbard model on ladders of 2, 4 and 6 legs with nearest $(t)$ and next-nearest $(t')$ neighbor hoppings by means of the density-matrix renormalization group algorithm. In particular, we analyze the stability of the Nagaoka state for several values of $t'$ when we vary the electron density $(\rho)$ from half-filling to the low-density limit. We build the two-dimensional phase diagram, where the fully spin-polarized and paramagnetic states prevail. We find that the inclusion of a non-frustrating next nearest neighbor hopping stabilizes the fully spin-polarized phase up until $|t'/t|=0.5$. Surprisingly, for this value of $t'$, the ground state is fully spin-polarized for almost any electron density $1 \gtrsim \rho \gtrsim 0$, connecting the Nagaoka state to itinerant ferromagnetism at low density. Also, we find that the previously found checkerboard insulator phase at $t'=0$ and $\rho=0.75$ is unstable against $t'$. 
\end{abstract}

\maketitle 

\section{Introduction} 

The Hubbard model was first introduced in 1963 to explain itinerant ferromagnetism in transition metals\cite{gutzwiller63}$^-$\cite{hubbard63}. Even though it was built as the simplest model capable of describing the behavior of correlated materials\cite{quintanilla09}, it has been proven to be useful in the interpretation of a wide variety of phenomena ranging from metal-insulator transitions\cite{hubbard64}$^,$\cite{imada98} to high-temperature superconductivity\cite{lee06}. In the past years this model has been brought back into focus due to its applicability in the description of ultracold atoms in optical lattices\cite{jaksch98}$^-$\cite{jordens08}.

Despite being a rather simple model when written down, most of the progress in the understanding of the Hubbard model has been made numerically, either by exact diagonalization of small clusters, mean field approaches or by using more sophisticated numerical techniques such as quantum Monte Carlo (QMC) or density-matrix renormalization group (DMRG). Only a few mathematically rigorous results regarding this model exist to date\cite{tasaki98}$^,$\cite{li14}. The Nagaoka's theorem\cite{nagaoka66}$^,$\cite{tasaki89} is one of these few exact results, making it an important and solid starting point to study the phase diagram of the Hubbard model. It states that, when the system has one electron less than half-filling, $U\to\infty$, and the lattice satisfies certain connectivity conditions, the ground state of the system is a fully spin-polarized ferromagnetic state (FSP) and it is unique apart from the trivial spin degeneracy. These connectivity conditions require that the smallest loop must be no longer than 4 sites and the kinetic energy of the hole motion around this loop must not be frustrated. In the subsequent years to the Nagaoka's theorem, a lot of effort was put into trying to widen this isolated point of the phase diagram by relaxing some of the requirements of this theorem. For example, very recently, the condition regarding the loop size has been extended to cover larger loops\cite{bobrow18}, proving that the two-dimensional honeycomb lattice (loop size $= 6$) also has a FSP ground state. 

When the accumulated sign of the hoppings along the minimum loop is negative, the kinetic energy of the hole motion is frustrated and the Nagaoka's theorem is no longer valid. Nevertheless, it has been shown that the FSP state can survive in the presence of small enough values of frustrating hoppings\cite{lisandrini17}. On the other hand, high frustration can lead to an antiferromagnetic N\'eel order, as it happens in the isotropic triangular lattice with positive hoppings\cite{haerter05}. Surprisingly enough, in this case, the antiferromagnetic ground state is classical and has the maximum staggered magnetization possible\cite{lisandrini17}$^,$\cite{sposetti14}. This kind of classical antiferromagnetic state has also been found in the square lattice with a frustrating next-nearest neighbor hopping\cite{sposetti14}.

Also, it is worth mentioning that the Nagaoka's theorem is only valid for finite lattices, where the one hole away from half-filling condition makes sense. As we get closer to the thermodynamic limit, clearly, this condition means that the electron density tends to half-filling. Because of this, there has been a long-standing question regarding the existence of the FSP phase in the thermodynamic limit at finite hole-doping. The Hubbard model on the square lattice with $U\to\infty$ and varying electron density is the easiest model where this problem can be studied. Early calculations showed that the FSP phase was unstable against hole doping beyond the Nagaoka's theorem conditions\cite{doucot89}$^,$\cite{shastry90}, although  more recent ones obtain a critical hole density in the thermodynamic limit around $\rho_c \approx 0.8$\cite{carleo11}$^-$\cite{becca01}. Nonetheless, there is still much to uncover, as other recently published results suggest that two holes away from half filling the existence of the FSP state depends strongly on the boundary conditions and the sizes of the finite lattices\cite{ivantsov17}. This shows that the mechanisms responsible for stabilizing the FSP state away from the Nagaoka's theorem conditions are not yet fully understood. 

DMRG calculations\cite{liu12} show that, also in the square lattice, new interesting phases appear below the critical electron density, $\rho_c = 0.8$. For example, a commensurate checkerboard insulator state emerges at $\rho = 0.75$, and leads to a phase separation region between $\rho=0.8$ and $0.75$; below this point the system behaves as a paramagnet. Another interesting study\cite{park08} performed with dynamical mean-field theory (DMFT) shows that the inclusion of a next-nearest neighbor hopping, whenever not frustrating the loop conditions within the Nagaoka's theorem, can stabilize ferromagnetic phases for smaller values of $\rho$ (reducing the paramagnetic region). 

The weak to intermediate coupling regime (\textit{i.e.} the infinite-$U$ condition is not longer fulfilled) has been widely studied\cite{hanisch97}$^-$\cite{igoshev12} and it was found that for low densities and $|t'/t|=0.5$ there is a fully spin-polarized phase. 
The presence of this low-density ferromagnetism even at relatively small values of $U$ is due to the strong particle-hole asymmetry and the van Hove singularity near the bottom of the band.

In this paper, we study in more detail the problem of the stability of the Nagaoka ferromagnetism (FSP state) against hole doping, combined with the inclusion of a non-frustrating next-nearest neighbor hopping on the square lattice by means of numerical calculations using DMRG. We also follow the evolution and stability of the checkerboard insulator previously mentioned along with the phase separation.

This paper is organized as follows: in Sec. \ref{sec2} we introduce the Hamiltonian and describe the details of the methods employed in the rest of the paper. In Sec. \ref{sec3} we show and discuss the results for two- to six-leg ladders and build the two-dimensional (2D) phase diagram. In Sec. \ref{sec4} we present the conclusions.

\section{Model and methods}
\label{sec2}
\begin{figure}[t]
\begin{center}
\includegraphics*[width=0.40\textwidth]{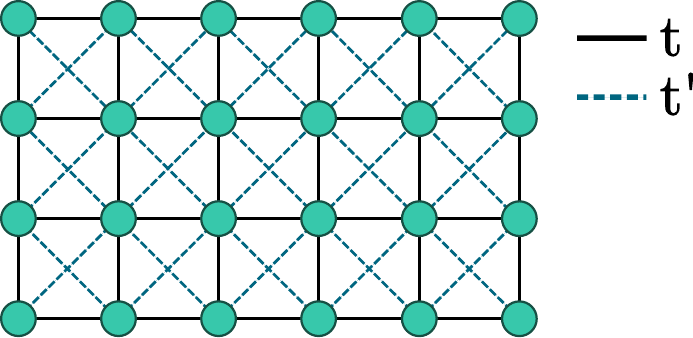}
\caption{(color online) $4 \times 6$ lattice with open boundaries conditions. In solid black lines $t$ is the nearest neighbor hopping integral and in dashed blue lines $t'$ is the next-nearest neighbor one.}
\label{fig1}
\end{center}
\end{figure}

The object of our present study is the Hubbard model, which we can write down as
\begin{equation}
\mathcal{H} = -\sum_{ ij  \sigma} t_{ij} \left[\hat{c}^\dagger_{i\sigma} \hat{c}_{j\sigma} + \text{h.c.} \right] + U \sum_i \hat{n}_{i\uparrow} \hat{n}_{i\downarrow},
\end{equation}
\noindent where $t_{ij}$ is the hopping integral between sites $i$ and $j$ and $U$ is the on-site repulsion. We take the repulsion between electrons $U \rightarrow \infty$ to remain within the Nagaoka's theorem conditions; this means that we can never have two electrons in the same site. Taking into account only two different hopping terms, we can write our implementation Hamiltonian as
\begin{equation}
\mathcal{H} = - t\sum_{ \langle ij \rangle  \sigma} \left[\tilde{c}^\dagger_{i\sigma} \tilde{c}_{j\sigma} + \text{h.c.} \right]+ t' \sum_{ \langle \langle ij \rangle \rangle \sigma}  \left[\tilde{c}^\dagger_{i\sigma} \tilde{c}_{j\sigma} + \text{h.c.} \right],
\end{equation}
\noindent where $t$ connects the nearest neighbors on a square lattice and $t'$ the next-nearest neighbors (see figure \ref{fig1}). The presence of a hopping $t'\neq 0$ makes the lattice no longer bipartite, breaking the particle-hole symmetry. Given the importance of the sign of the interaction in the Hamiltonian, it is worth mentioning that we have chosen it differently for $t$ and $t'$ for later convenience. The new operators $\tilde{c}^\dagger_{i\sigma}=\hat{c}^\dagger_{i,\sigma} (1-\hat{n}_{i,\bar{\sigma}})$ ensure the exclusion of the doubly-occupied states imposed by the infinite-$U$ condition; and its new commutation relations are responsible for the complications in diagonalizing the Hamiltonian. 

Also, to ensure the validity of the Nagaoka's theorem connectivity condition we need to check that the accumulated sign around the smallest loop is positive, meaning that
\begin{equation}
\text{sign}(t')(\text{sign}(-t))^2 = 1
\end{equation}
\noindent where we can see that the sign of $t$ is irrelevant. So $t'$ has to be positive to fulfill the connectivity condition and we will take $t=1$ as the energy unit from now on. Taking $t'>0$, we can be sure that the ground state of the system will always be the FSP state at one electron less than half-filling.

Having guaranteed this starting point, it is up to us to study in further detail the effect of the inclusion of next-nearest neighbor hopping terms $t'$ over the existence and stability of the FSP state upon further hole-doping. To do so, we use the density-matrix renormalization group algorithm\cite{white92} based on the matrix product state (MPS) representation\cite{schollwock11} contained in the ALPS libraries\cite{bauer11}$^,$\cite{dolfi14}. We solve ladders with fully open boundary conditions for $L_y=$2, 4, 6 (legs) and several $L_x$ (rungs) when possible. We vary $t'$ and the electron density $\rho$ to map the phase diagram, using between $m=2000$ and $9000$ DMRG states to ensure the convergence of our results. 

The first of our aims is to determine the critical value of $\rho_c$ up to which the FSP state survives. In order to accomplish this task, we need a reliable signature to help us determine whether this state is, or not, the ground state of the system for a given hopping $t'$ and electron density $\rho$. There are two complementary methods we can use that will also help us characterize the rest of the phase diagram. One is to determine the magnetization of the system through the spin structure factor
\begin{equation}
S({\bf k}) = \frac{1}{N} \sum_{ij} \langle {\bf S}_i {\bf S}_j \rangle e^{-i{\bf k}({\bf r}_i - {\bf r}_j)},
\end{equation}
where we can take ${\bf k}$ as a continuous variable given the open boundary condition. Obtaining the spin structure factor of the system gives out more information about the magnetic ordering, but it relies on the calculation of the spin correlations, which are considerably less accurate than the energy calculations. The other alternative depends on calculating the ferromagnetic magnetization of a system solely through energy calculations. This allows us to calculate the total spin of the ground state by exploiting the degeneracy of a ferromagnetic ground state. To do so, we calculate the lowest energy for all subspaces with different quantum number $S_z$, and find $S_z^\text{max}$ such that $E(S_z=0) = ... =E(S_z=S_z^\text{max}) < E(S_z=S_z^\text{max}+1).$ Then, the normalized ferromagnetic magnetization value for a given electron density can be calculated as $M=2S_z^\text{max}/N_e$. In the case of the FSP state, this magnetization gives out $M=1$ and in the case of the paramagnetic state, $M=0$. This method is expected to be more accurate but more time-consuming as we have to compute the minimum energy state for several subspaces. Nevertheless, there is an issue with the given $M$ formula: when there is an odd number of electrons in the system, the minimum value of the spin projection $S_z$ is $0.5$ (and not $0$), meaning that the minimum value of $M$ is not zero. To solve this, in these cases we subtract to the $S_z^{max}$ one half, and one electron to the $N_e$ to be able to represent the FSP and paramagnetic limits, $M=1$ and $M=0$. The corresponding formula of the magnetization for the odd number of electrons then results in $M=(2S_z^\text{max}-1)/(N_e-1)$.

\section{Results}
\label{sec3}
\subsection{Two-leg ladders}

We start by studying the simpler 2-leg ladder systems as we expect that the main characteristics of the phase diagram do not change much upon adding more legs. To cover the phase diagram we use $2 \times 20$ and $2 \times 30$ ladders and take $t'$ from $0.0$ to $0.5$ at $0.1$ intervals. Also, we briefly comment results for $t' > 0.5$ up to $t' = 1.0$. For each chosen value of $t'$ we calculate the ground state energy within all $S_z$ subspaces for all electron densities below half-filling. This allows us to compute the magnetization curves for all our $t'$ values as a function of the electron density $\rho$. 

\begin{figure}[t!]
\begin{center}
\includegraphics*[width=0.4\textwidth]{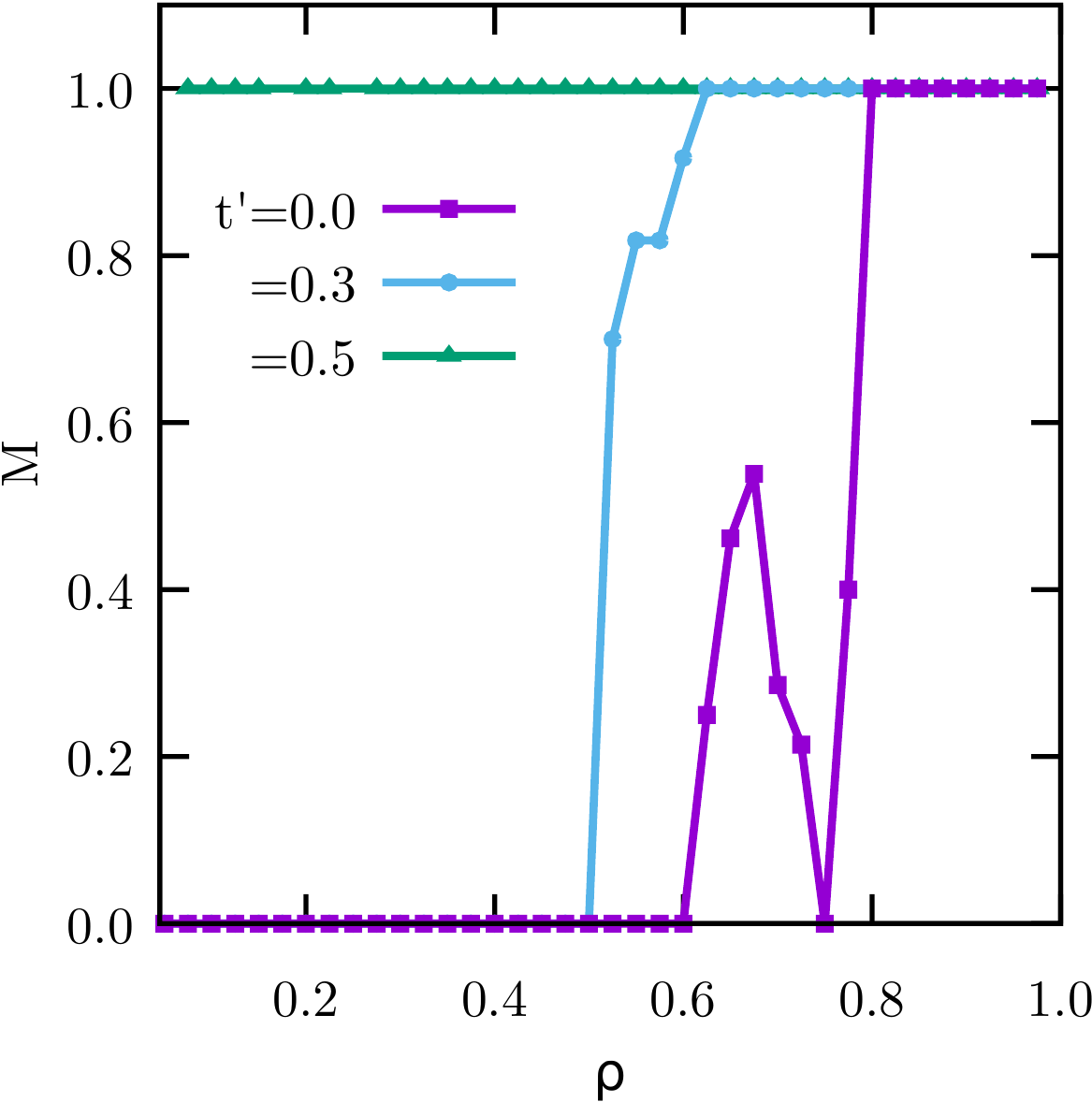}
\caption{(color online) Magnetization value in $2 \times 20$ ladders with open boundaries conditions for three different values of $t'$. $M=1$ is the fully spin-polarized state and $M=0$ is the paramagnetic state. In purple squares the limit $t'=0$ (square lattice), in blue circles an intermediate case $t'=0.3$, and in green triangles the limit case of $t'=0.5$.}
\label{fig2}
\end{center}
\end{figure}

\begin{figure}[h!]
\begin{center}
\includegraphics*[width=0.48\textwidth]{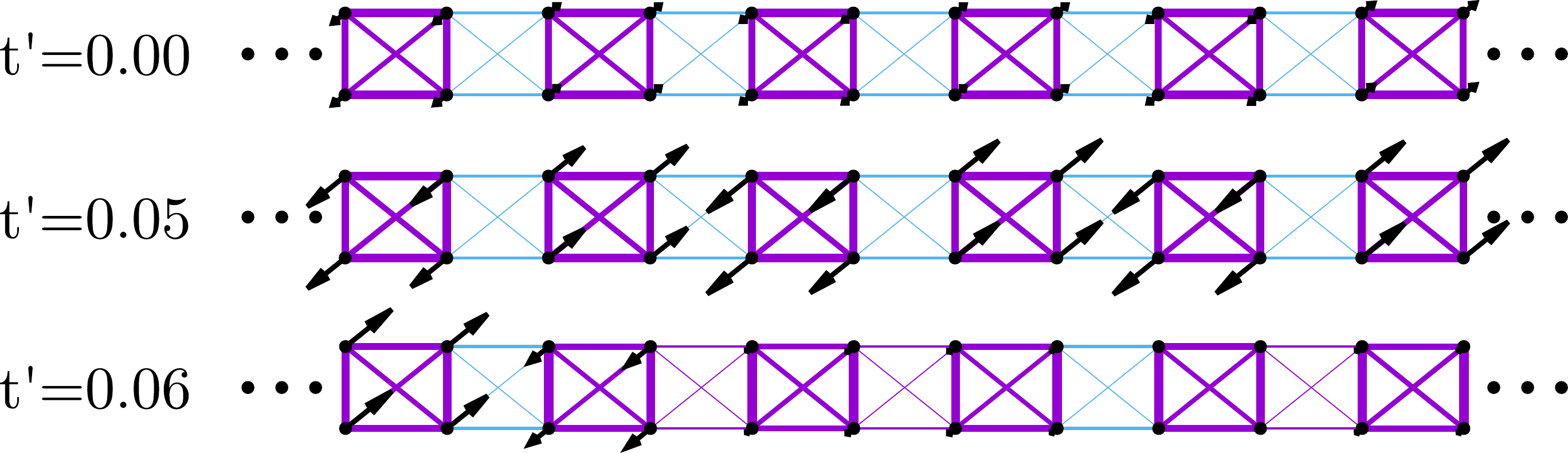}
\caption{(color online) Nearest and next-nearest neighbor correlations in the central part of the $2\times20$ lattice for several values of $t'$. The thickness of the lines of the lines varies as the third power of the bond density $B_{ij}=\sum_\sigma\langle [\tilde{c}^\dagger_{i\sigma} \tilde{c}_{j\sigma} + \text{h.c.} ] \rangle$, while the color indicates the sign of the total spin correlations. The size of the arrows indicates the magnitude of the local value of $S^z_i$.}
\label{fig3}
\end{center}
\end{figure}

In the square ladder ($t'=0.0$), the FSP state is present at high densities and survives up to $\rho_c =0.8$, as can be seen by the classical magnetization $M=1$ region in figure \ref{fig2}. This value was previously obtained in 2-leg ladder systems\cite{krivnov98}$^,$\cite{liu12} and in the 2D limit\cite{carleo11}$^,$\cite{liu12}, proving that it does not scale with the number of legs. Below this critical value of the electron density, the magnetization $M$ is lowered up until $\rho = 0.75$, where it reaches $M=0$. At $\rho=0.75$ lies the checkerboard insulator; this phase consists on plaquettes of four sites and three electrons (each plaquette has the same electronic density as the lattice) in a FSP state, but building an antiferromagnetic order between plaquettes (see figure \ref{fig3}). This exotic state can be seen through the spin structure factor as a set of two broad peaks centered at $\pm \frac{\pi}{2}$ (shown in next section for larger ladders). Between this phase and the FSP there is a phase separation region that arises as a combination of the FSP state with $\rho = 0.8$ and the checkerboard insulator. This phase separation region can be seen in the spin structure factor as a combination of the peaks that belong to each of the phases, lowered and broadened by the mixture (also shown in the next section for larger ladders). This phase can also be characterized by the uneven charge distribution \cite{liu12}, where a certain part of the system has the same density  $n=0.75$ and order as the checkerboard insulator whereas the rest of the system has ferromagnetic order with $n=0.8$. Between $\rho=0.75$ and $\rho=0.6$ there are intermediate phases with non-zero magnetization which disappear in wider ladders. Below $\rho=0.6$ the ground state is paramagnetic, and it is signaled by the null ferromagnetic magnetization. These results, obtained for $2 \times 20$ and $2 \times 30$ lattices, are in complete agreement with the previous DMRG study by Liu {\it et al.}\cite{liu12}; and they provide us a good benchmark to start analyzing the effect of introducing the next-nearest neighbor hopping $t'$.

When $t'$ is turned on, we find that the FSP state region starts growing and $\rho_c$ is lowered, as can be seen in figure \ref{fig2}. For example, when $t'$ increases to $0.30$, the critical value of the electron density for which the FSP can be found moves down to $\rho_c = 0.62$. This enhancement of the stability of the FSP phase when including $t'$ has also been found using DMFT in 2D\cite{park08}, where they report that at $t'=0.1$ the critical value of the density is $\rho_c=0.705$. For the same value of $t'$, we obtain $\rho_c=0.775$. The difference in our results may be a signal that $\rho_c$ depends on the number of legs when approaching to the 2D limit for $t'\neq 0$ (unlike what happens for $t'=0.0$).
To understand why the $t'$ stabilizes the FSP state, Park {\it et al.}\cite{park08} have solved the four-site plaquette with three electrons. They have shown that the existence of $t'$ lowers more the FSP state energy than the low-spin state energy because of the quantum interference of different hole paths. As a consequence, the gap between these two states increases with $t'$. 

It is noteworthy that, if the value of the next-nearest neighbor hopping is half of the hopping on the square lattice, $t'=0.5$, the FSP state survives for every value of electron density; that is $\rho_c \rightarrow 0$ (see green triangles in figure \ref{fig2}).
This is a remarkable result, because it connects the Nagaoka ferromagnetic phase (an exact result near half-filling from the infinite-$U$ limit) with the FSP state found at low density with relatively small $U$.
The latter phase was studied among others by Taniguchi {\it et al.}\cite{taniguchi05} and they found this ferromagnetic phase for a rather small $U$ in the low-density limit around $t'\simeq 0.5$, where the Fermi energy is close to the van Hove singularity. They have also shown that with increasing $U$ (until $U=5$) the FSP phase is the ground-state of the system for a wider regime of values of $t'$ and $\rho$.

When increasing $t'$ above $0.5$, the FSP state region starts going back, giving in to the paramagnetic phase. This is a consequence of the magnetic behavior of the system when $t' \to \infty$, in this case the ladder splits in two independent chains where the ground states is a paramagnet\cite{ogata90}.
Also, for every value of $t'\neq 0$ we find intermediate phases with interpolating ferromagnetic magnetization between the FSP and paramagnetic regions, but we expect them to shrink and disappear when adding more legs and going closer to the 2D limit. An important difference with the $t'=0.0$ case is that there is no checkerboard insulator from $t'=0.1$ on. This also means that there is no phase separation.

\begin{figure}[t!]
\begin{center}
\includegraphics*[width=0.4\textwidth]{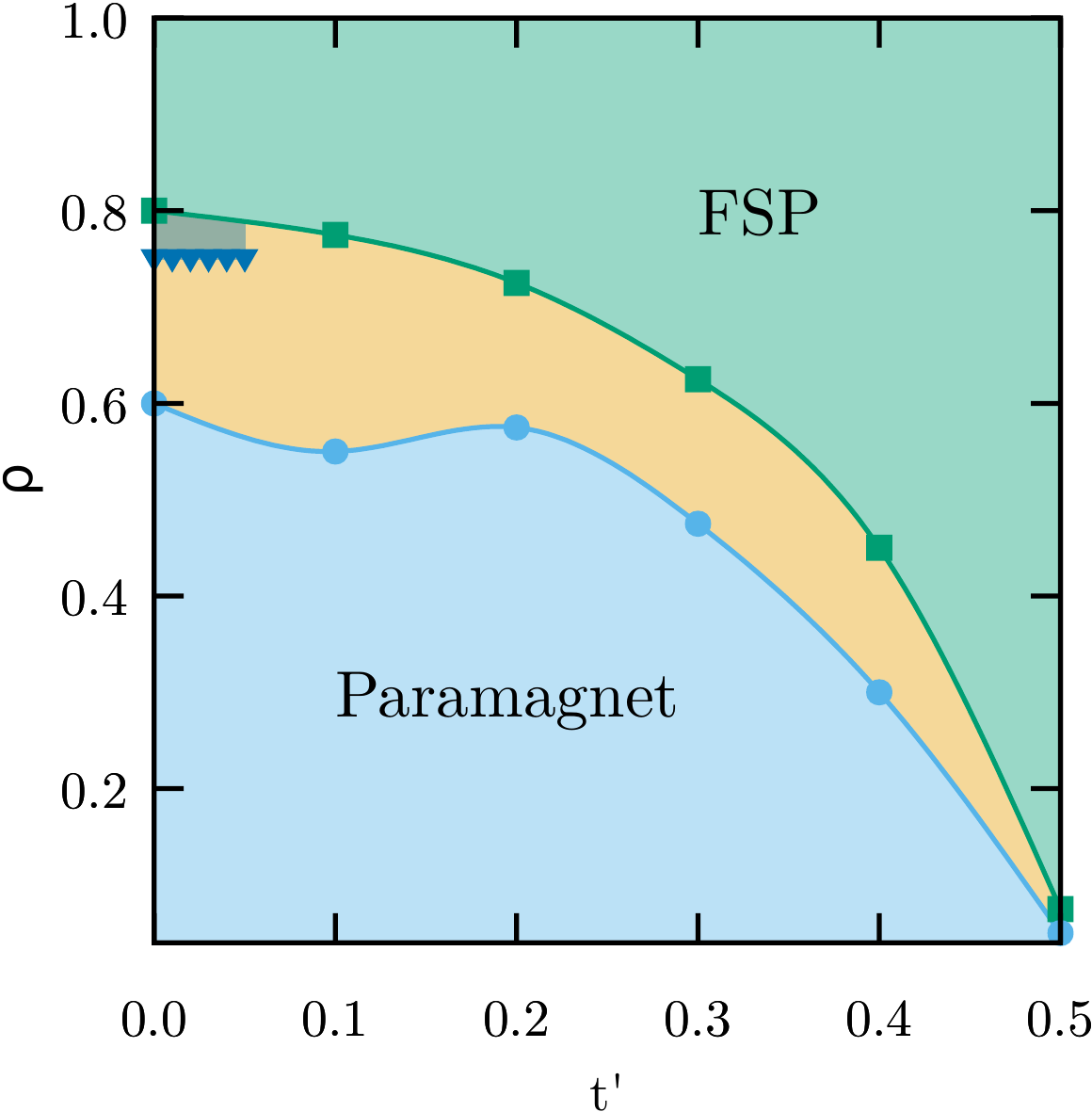}
\caption{(color online) Phase diagram for the $2 \times 20$ lattice with open boundaries conditions as a function of the next-nearest neighbor hopping $t'$ and electron density $\rho$. In green squares the boundary of the fully spin-polarized ferromagnetic state and in blue circles the boundary of the paramagnetic state. The dark blue triangles represent the points where checkerboard insulator is found. Points are connected by smooth spline fits.}
\label{fig4}
\end{center}
\end{figure}

Given that the checkerboard insulator phase is absent even at $t'=0.1$, we decided to follow its evolution more closely. Taking a fixed electron density $\rho=0.75$, we made runs varying $t'$ at $0.01$ intervals. In figure \ref{fig3} we plot the results for $t'=0.00$, $0.05$, and $0.06$. Clearly, the four-plaquette structure can be seen up until $t'_c=0.05$, but as soon as $t'$ is increased this structure disappears and the ferromagnetic magnetization rises.

Regarding the validity of these results for longer ladders, it can be seen in figure \ref{fig5} that the results do not change upon adding more rungs. The $2\times20$ lattice already is a good representation of the 2-leg thermodynamic limit for every value of $t'$. The previous discussion on the magnetization can be summarized into the phase diagram shown in figure \ref{fig4}. The FSP phase is painted in green, being the green squares the last points for which we find this kind of state, coming down from the one electron less than half-filling condition of the Nagaoka's theorem. In blue we show the paramagnetic phase, being the blue circles the last points for which we find a paramagnetic phase, with $M=0$, when increasing the electron density from zero. In both of these cases it is usually easy to determine, by studying the dependence of the energy with the value of $S_z$, if the ground state lies only on the $S_z=0$ subspace or if it can be found in all subspaces up to $S_z^\text{max}$. The blue triangles in line at $\rho =0.75$ represent the points of the phase diagram for which we find the checkerboard insulator phase. In the middle of this phase and the FSP, we shadowed the area in which the phase separation exists. The yellow region corresponds to the intermediate phases with intermediate ferromagnetic magnetization. Closer to the FSP they present clear ferromagnetic peaks in the spin structure factor that evolve quickly into the paramagnetic state. Here, the tendencies of the energy as a function of $S_z$ make it more complicated to identify the exact value of $S_z^\text{max}$, and therefore to analyze the properties of the ground state in this region. As mentioned above, increasing $t'$ above $0.5$ causes the FSP region to shrink again (and the paramagnetic one to grow). 

With respect to the charge distribution, we find that both the paramagnetic and the fully spin-polarized ferromagnetic phases show almost homogeneous distribution. On the other hand, the intermediate phases show a minor variation of the charge distribution along the ladder which, however, does not resemble that of a separation of phases where only two well distinctive densities appear. 

\subsection{Four- and six-leg ladders}

To shed some light on the 2D behavior of the system we extended our calculations to 4- and 6-leg ladders. Leaning on our results for the 2-leg ladders, we calculate the magnetization value in $4 \times 10$, $4 \times 12$ and $6 \times 8$ lattices around the transition points for all the same values of $t'$ as in the 2-leg case to see how they scale. Also, we calculate the spin structure factor and other correlations. Away from the transition points the convergence of the results behaves very well even for the bigger lattices. But, close to transition, the computational effort increases and it becomes difficult to determine precisely the $S_z^\text{max}$ of the ground state of the system. This is the main reason why we had to limit the number of rungs and legs used in our lattices. Previous studies already show that, even at $t'=0.0$, the amount of DMRG states needed to obtain accurate results in big lattices is huge\cite{liu12}. 

\begin{figure}[t!]
\begin{center}
\includegraphics*[width=0.4\textwidth]{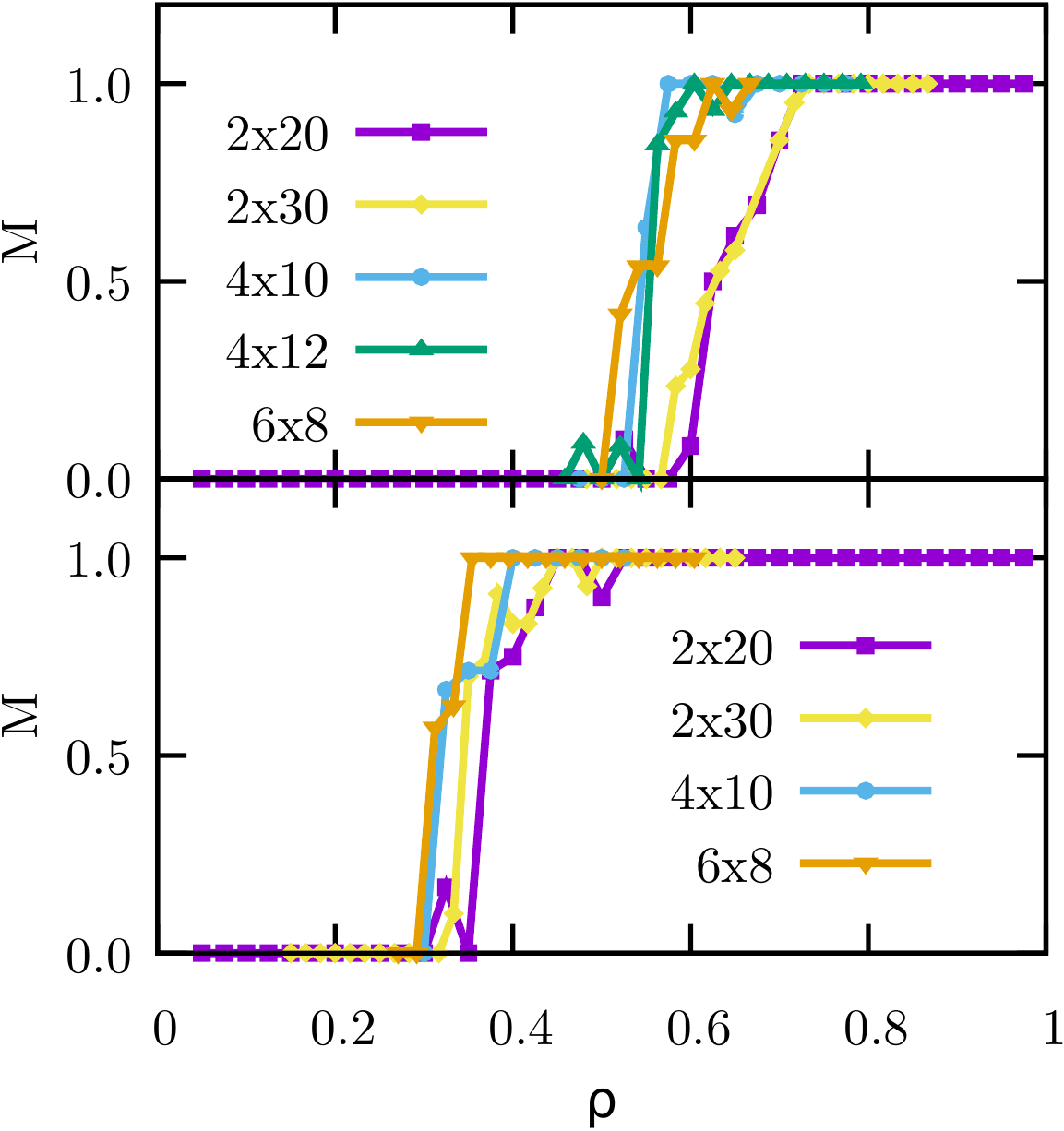}
\caption{(color online) Magnetization value at $t'=0.2$ (top panel) and $t'=0.4$ (lower panel) as a function of $\rho$ for several lattice sizes: $2 \times 20$ (purple squares), $2 \times 30$ (yellow diamond), $4 \times 10$ (blue circles), $4\times12$ (green up triangles), and $6 \times 8$ (orange down triangles). }
\label{fig5}
\end{center}
\end{figure}

We show on figure \ref{fig5} the magnetization value $M$ for two arbitrary values of next-nearest neighbor hoppings $t'=0.2$ (top panel) and $t'=0.4$ (lower panel), and several lattice sizes. From these results we can see the qualitative behavior remains the same for all the lattices. For example, in the top panel we can see that the critical value $\rho_c$ seems to move a little in comparison with the 2-leg lattice, but we have roughly the same critical density for the 4- and 6-leg lattices. Also, it seems that the region corresponding to intermediate phases shrinks as the FSP phase region grows, making the transition to the paramagnetic phase more abrupt, also seen in the lower panel. We have observed that this behavior, shown for the $t'=0.2$ and $t'=0.4$ cases, holds for almost every hopping value below $t'=0.5$; $t'=0.0$ being the special case. At $t'=0.0$, the checkerboard insulator prevails and prevents the FSP state to take over and, instead, the intermediate phases become paramagnetic. On the other hand, when $t'=0.5$ there is already no place to move the critical electron density down, so the ground state remains FSP for all values of the electron densities below half-filling. For $t'=0.1$, we find $\rho_c = (0.715\pm 0.015)$, much closer to the value $\rho_c = 0.705$ obtained by DMFT\cite{park08}. For $t'=0.0$ both methods also agree, as $\rho_c=0.8$ in DMRG and $\rho_c=0.815$ in DMFT. These similarities between our DMRG results and DMFT seem to indicate that, around the phase transition, the non-local physics are unimportant. Nonetheless, this could be a mere coincidence and more research is needed to elucidate the source of this agreement. We can then conclude that the 4- and 6-leg lattices provide us enough information about the scaling of the transition points to build the 2D phase diagram of the system until $t'=0.5$, at least up to a certain small error.

\begin{figure}[t!]
\begin{center}
\includegraphics*[width=0.5\textwidth]{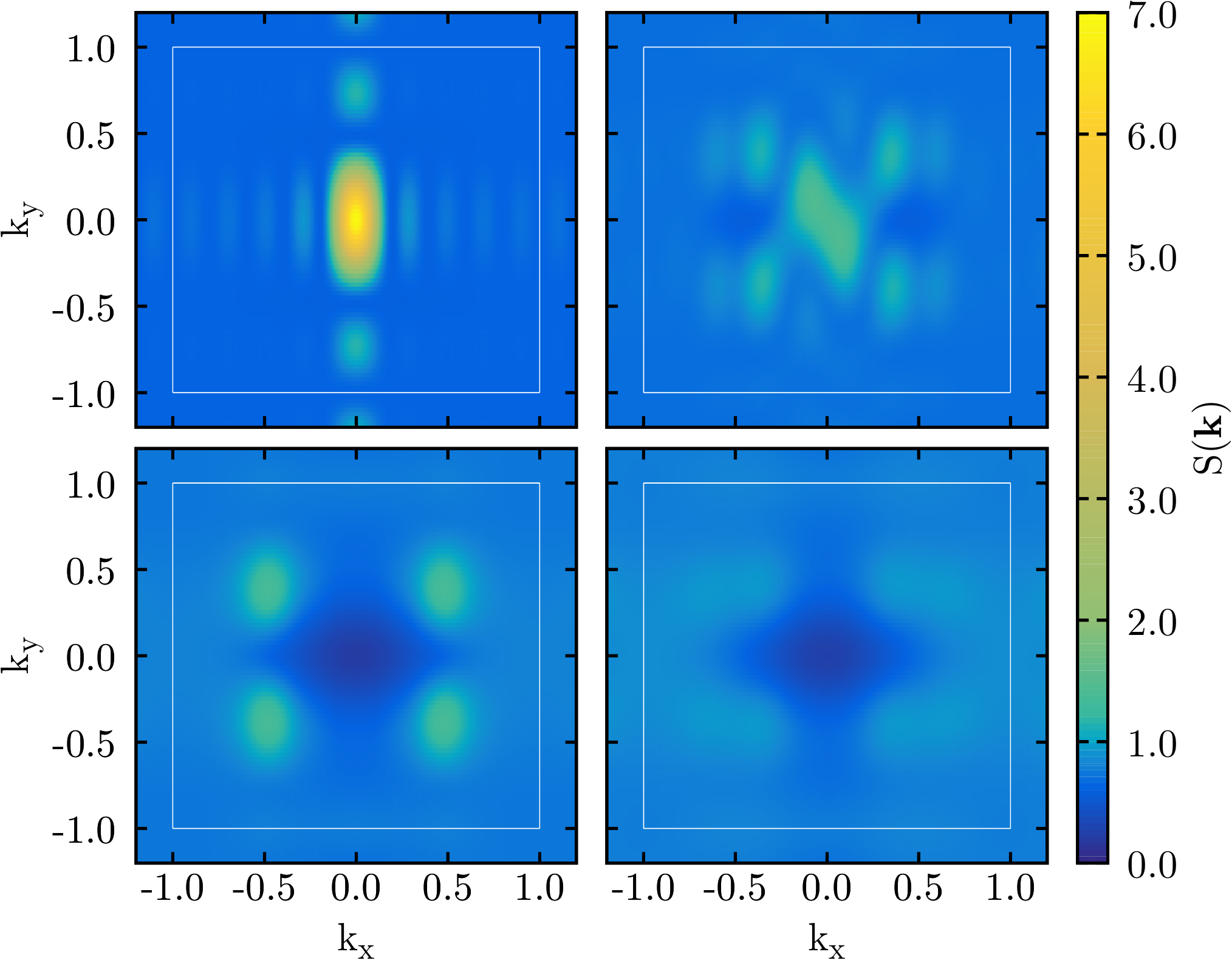}
\caption{(color online) Spin structure factor of the $4\times10$ lattice for $t'=0.0$ and several electron densities $\rho$: 0.8 (upper-left), 0.775 (upper-right), 0.75 (lower-left), and 0.725 (lower-right). The white square represents the first Brillouin zone, and $k_x$ and $k_y$ are in units of $\pi$.}
\label{fig6}
\end{center}
\end{figure}

Above $t'=0.5$ the picture is completely different. We found a more pronounced scaling of the critical densities and intermediate phases that made it not possible for us to extract valuable information. One possible reason for this to happen has to do with the large $t'$ limit. While in the 2-leg ladders the $t'\to \infty$ limit results in two isolated chains, in a 2D system it results in two sets of independent square lattices. The difference in these limits may be responsible for the behavior above $t'=0.5$. A more detailed study in wider ladders is needed and this is why we restrain our results to the region below $t'=0.5$.
 
In figure \ref{fig6} we show, for the $4\times10$ lattice, several calculations of the spin structure factors obtained at $t'=0.0$. In the upper-left panel we can see the last density for which the ground state of the system is FSP, $\rho = 0.8$. This phase is signaled by a large peak centered at ${\bf k} = {\bf 0}$ with certain asymmetry given by the lattice. In the lower-left panel we can see the signature of the checkerboard insulator, $\rho = 0.75$. The structure factor is composed by four peaks situated at ${\bf k} = \left(\pm \frac{\pi}{2},\pm \frac{\pi}{2}\right)$ that arise from the antiferromagnetic alternation of the plaquettes. Note that, in this case, the ferromagnetic magnetization is zero and $S({\bf k = 0})=0$, because the ground state exists only at $S_z=0$. In the upper-right panel we can see the phase separation at $\rho = 0.775$, signaled by a combination of the surrounding spin structure factors. It shows a low ferromagnetic peak and the checkerboard insulator weakened four-peak structure. Finally, in the lower-right panel we show the paramagnetic phase at $\rho = 0.725$ where no clear peak can be seen. It is important to be aware that, for this lattice size, all these states are one electron away from each other. The density step is $\frac{1}{40} = 0.025$.

\begin{figure}[t!]
\begin{center}
\includegraphics*[width=0.4\textwidth]{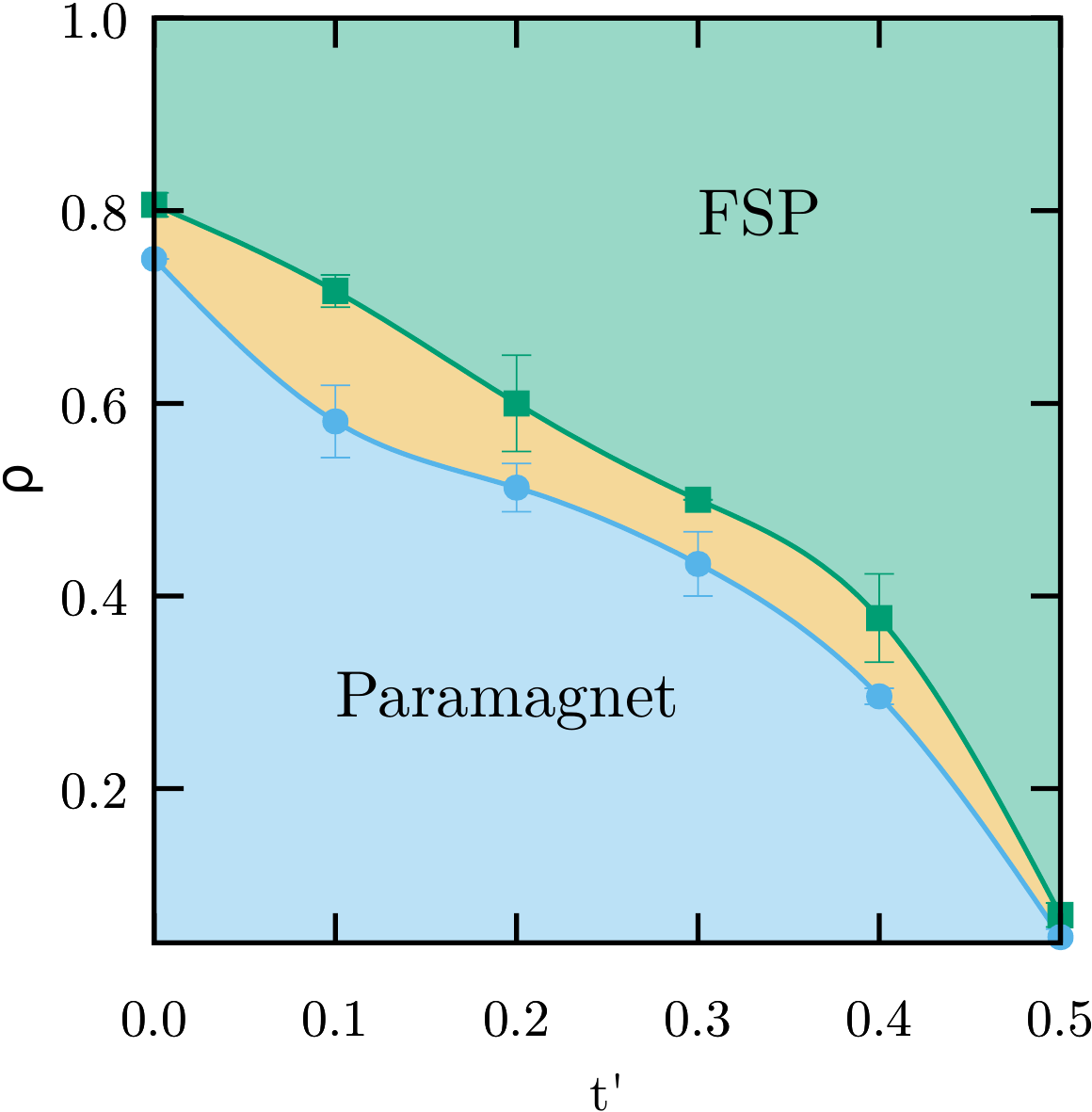}
\caption{(color online) 2D phase diagram of the system based upon results for 4- and 6-leg ladders. In green we show the FSP phase, in blue the paramagnetic one, and in yellow the intermediate region. The green squares and blue circles signal the end-points of the FSP and paramagnetic phases, respectively. Points are connected by smooth spline fits.}
\label{fig7}
\end{center}
\end{figure}

With all this gathered information, now we can compute the full phase diagram (shown in figure \ref{fig7}) and compare it with the 2-leg case. We have decided to take the transition points as the average of the ones in 4- and 6-leg ladders, with error equal as the difference in these results. As we expected before, the 2D phase diagram resembles the 2-leg ladders one, but with less area left for intermediate phases, specially around $t'=0.1$, and a certain growth of the FSP phase. In contrast with $t'=0$, the intermediate phases shrink but they do not disappear in the four- and six-leg ladders. For these ladders, the intermediate phase is ordered ferromagnetically (peak in the structure factor at $S({\bf k = 0})$) but its magnetization is not saturated.
The checkerboard insulator, even though we clearly find it in the 4-leg lattices (as can be seen in figure \ref{fig6}),  was much harder to find in the $6 \times 8$ lattice. We had to use $m=9000$ DMRG states and a small Zeeman field applied to pin this order; and check that the arising checkerboard insulator is not an excited state. Given that in the 4- and 6-leg lattices we observed that this phase vanishes when $t'\sim 0.02$ (much smaller than in the 2-leg case), we conjecture that the phase boundaries for the checkerboard insulator phase in the 2D limit should be rather small ($t'_c\sim 0.02$ at most). This is in consonance with the small spin gap of the checkerboard phase at $t'=0$, about $10^{-3}\, t$.

\section{Conclusions}
\label{sec4}
We have used density-matrix renormalization group to study the phase diagram of the infinite-U Hubbard model on square ladders with nearest and next-nearest neighbor hopping amplitudes $t$ and $t'$, respectively; and always in the absence of kinetic frustration.
 We have found that, for all ladder sizes, the presence of a non-frustrating next-nearest neighbor hopping amplitude stabilizes the fully spin-polarized phase. With increasing $t'$, the fully spin-polarized phase region grows in the phase diagram until $t'$ reaches a value of one half the nearest neighbor hopping amplitude. For this particular value ($t'=0.5$) the ground state of the system is always a fully spin-polarized state, regardless of the electron density chosen below half-filling. We have connected in the infinite-$U$ limit the fully spin-polarized state from the Nagaoka's theorem (valid for one hole over half-filling) with the low-density ferromagnet (also FSP) that arise due to the van Hove singularity in the bottom of the band. It would be interesting to further investigate the behavior of the single-particle spectral densities in the different phases, and in particular for $t'=0.5$ in the Nagaoka's theorem case and in the low-density regime, to unwrap the transition between different types of ferromagnetism. For $t'>0.5$ we need to explore wider ladders to uncover the 2D behavior, but our results indicates that beyond this point, the fully spin-polarized phase region starts to shrink as $t'$ increases.

With regards to the intermediate phases, we conjecture that the previously found checkerboard insulator phase only survives for small values of $t'$ in the thermodynamic limit. This may be a consequence of the proximity to the fully spin-polarized state. The checkerboard insulator phase can only exist at $\rho=0.75$, where one hole lives in each four-site plaquette. But, $\rho_c$ quickly goes below 0.75 when $t'$ is included and the fully spin-polarized state prevails over the checkerboard insulator. 
Moving away from the $t'= 0$ case, the intermediate phases are less interesting and have generally a ferromagnetic behavior which seems to connect continuously the FSP phase with the paramagnetic phase. In this region it is more difficult to analyze the spin and charge behavior so we cannot ensure the size of this region or characterize the nature of the transition in the thermodynamic limit.

Also, we expect these results to be of interest and to contribute to the on-going research in optical lattices, which is where these models with large repulsions within particles can be experimentally realized. 

\section{Acknowledgment}
The authors would like to thank L.O. Manuel for fruitful discussions.
This work was supported by CONICET (Argentina) under grant Nro. 0364 (PIP2015).


\begin{thebibliography}{99}

\bibitem{gutzwiller63} M. C. Gutzwiller,
Phys. Rev. Lett. {\bf 10}, 159 (1963).

\bibitem{kanamori63} J. Kanamori,
Prog. Theor. Phys. {\bf 30}, 275 (1963).

\bibitem{hubbard63} J. Hubbard,
Proc. R. Soc. A {\bf 276}, 238 (1963).

\bibitem{quintanilla09} J. Quintanilla and C. Hooley,
Phys. World {\bf 22}, 32 (2009).

\bibitem{hubbard64} J. Hubbard,
Proc. R. Soc. A {\bf 281}, 401 (1964).

\bibitem{imada98} M. Imada, A. Fujimori, and Y. Tokura,
Rev. Mod. Phys. {\bf 70}, 1039 (1998).

\bibitem{lee06} P. A. Lee, N. Nagaosa, and X.-G. Wen,
Rev. Mod. Phys. {\bf 78}, 17 (2006).

\bibitem{jaksch98} D. Jaksch, C. Bruder, J. I. Cirac, C. W. Gardiner, and P. Zoller,
Phys. Rev. Lett. {\bf 81}, 3108 (1998).

\bibitem{greiner02} M. Greiner, O. Mandel, T. Esslinger, T. W. H\"ansch, and I. Bloch,
Nature (London) {\bf 415}, 39 (2002).

\bibitem{jordens08} R. J\"ordens, N. Strohmaier, K. G\"unter, H. Moritz, and T. Esslinger,
Nature (London) {\bf 455}, 204 (2008).

\bibitem{tasaki98} H. Tasaki,
J. Phys. Condens. Matter {\bf 10}, 4353 (1998).

\bibitem{li14} Y. Li, E. H. Lieb, and C. Wu,
Phys. Rev. Lett. {\bf 112}, 217201 (2014).

\bibitem{nagaoka66} Y. Nagaoka, 
Phys. Rev. {\bf 147}, 392 (1966).

\bibitem{tasaki89} H. Tasaki,
Phys. Rev. B {\bf 40}, 9192 (1989).

\bibitem{bobrow18} E. Bobrow, K. Stubis, and Y. Li,
Phys. Rev. B {\bf 98}, 180101(R) (2018).

\bibitem{lisandrini17} F. T. Lisandrini, B. Bravo, A. E. Trumper, L. O. Manuel, and C. J. Gazza,
Phys. Rev. B {\bf 95}, 195103 (2017).

\bibitem{haerter05} J. O. Haerter and B. S. Shastry,
Phys. Rev. Lett. {\bf 95}, 087202 (2005).

\bibitem{sposetti14} C. N. Sposetti, B. Bravo, A. E. Trumper, C. J. Gazza, and L. O. Manuel,
Phys. Rev. Lett. {\bf 112}, 187204 (2014).

\bibitem{doucot89} B. Doucot and X. G. Wen,
Phys. Rev. B {\bf 40}, 2719(R) (1989).

\bibitem{shastry90} B. S. Shastry, H. R. Krishnamurthy, and P. W. Anderson,
Phys. Rev. B {\bf 41}, 2375 (1990).

\bibitem{carleo11} G. Carleo, S. Moroni, F. Becca, and S. Baroni,
Phys. Rev. B {\bf 83}, 060411(R) (2011).

\bibitem{krivnov98} V. Ya. Krivnov, A. A. Ovchinnikov,
Phys. Lett. A {\bf 248}, 453-456 (1998).

\bibitem{liu12} L. Liu, H. Yao, E. Berg, S. R. White, and S. A. Kivelson, 
Phys. Rev. Lett. {\bf 108}, 126406 (2012).

\bibitem{becca01} F. Becca and S. Sorella,
Phys. Rev. Lett. {\bf 86}, 3396 (2001).

\bibitem{ivantsov17} I. Ivantsov, A. Ferraz, and E. Kochetov,
Phys. Rev. B {\bf 95}, 155115 (2017).

\bibitem{park08} H. Park, K. Haule, C. A. Marianetti, and G. Kotliar,
Phys. Rev. B {\bf 77}, 035107 (2008).

\bibitem{hanisch97} T. Hanisch, G. S. Uhrig, and E. M\"uller-Hartmann,
Phys. Rev. B {\bf 56}, 13960 (1997).

\bibitem{hlubina97} R. Hlubina, S. Sorella, and F. Guinea,
Phys. Rev. Lett. {\bf 78}, 1343 (1997).

\bibitem{hlubina99} R. Hlubina,
Phys. Rev. B {\bf 59}, 9600 (1999).

\bibitem{arrachea00} L. Arrachea,
Phys. Rev. B {\bf 62}, 10033 (2000).

\bibitem{taniguchi05} H. Taniguchi, Y. Morita, and Y. Hatsugai,
Phys. Rev. B {\bf 71}, 134417 (2005).

\bibitem{igoshev12} P. A. Igoshev, A. V. Zarubin, A. A. Katanin, and V. Yu. Irkhin,
J. Magn. Magn. Mater. {\bf 324}, 3601 (2012).

\bibitem{white92} S. R. White,
Phys. Rev. Lett {\bf 69}, 2863 (1992).

\bibitem{schollwock11} U. Schollw\"ock,
Ann. Phys. {\bf 326}, 96 (2011).

\bibitem{bauer11} B. Bauer, L. D. Carr, H. G. Evertz, A. Feiguin, J. Freire, S. Fuchs, L. Gamper, J. Gukelberger, E. Gull, S. Guertler, A. Hehn, R. Igarashi, S. V. Isakov, D. Koop, P. N. Ma, P. Mates, H. Matsuo, O. Parcollet, G. Pawowski, J. D. Picon, L. Pollet, E. Santos, V. W. Scarola, U. Schollwck, C. Silva, B. Surer, S. Todo, S. Trebst, M. Troyer, M. L. Wall, P. Werner, and S. Wessel,
J. Stat. Mech. (2011) P05001.

\bibitem{dolfi14} M. Dolfi, B. Bauer, S. Keller, A. Kosenkov, T. Ewart, A. Kantian, T. Giamarchi, and M. Troyer,
Comput. Phys. Commun. {\bf 185}, 3430 (2014).

\bibitem{ogata90} M. Ogata and H. Shiba, 
Phys. Rev. B {\bf 41}, 2326 (1990).

\end{thebibliography}
\end{document}